\documentclass[twocolumn,showpacs]{revtex4}
\usepackage{epsfig}
\usepackage{graphicx}%
\usepackage{dcolumn}
\usepackage{amsmath}

\makeatletter
\def\btt#1{\texttt{\@backslashchar#1}}%
\DeclareRobustCommand\bblash{\btt{\@backslashchar}}%
\makeatother

\begin{document}

\title{Born-Infeld-type phantom on the brane world}

\author{Dao-jun Liu}
\author{Xin-zhou Li}\email{kychz@shtu.edu.cn}
\affiliation{Shanghai United Center for Astrophysics(SUCA),
Shanghai Normal University, 100 Guilin Road, Shanghai 200234,China
}%

\date{\today}

\begin{abstract}
ABSTRACT: We study the evolution of Born-Infeld-type phantom in
the second Randall-Sundrum brane scenario, and find that there
exists attractor solution for the potential with a maximum, which
implies a cosmological constant at the late time. Especially, we
discuss the BI model of constant potential without and with dust
matter. In the weak tension limit of the brane, we obtain an exact
solution for the BI phantom and scale factor and show that there
is no big rip during the evolution of the brane.
\end{abstract}

 \maketitle

The result of the recent astronomical observations
\cite{newobservation}, including WMAP \cite{WMAP}, indicate that
about seventy percent of the energy density in our universe is in
the form of dark energy that has negative pressure and can drive
the accelerating expansion of the universe. Many candidates for
dark energy have been proposed so far to fit the current
observations. Among these models, the most important ones are
cosmological constant and a time varying scalar field evolving in
a specific potential, referred to as "quintessence"
\cite{steinhardt}. The major difference among these models are
that they predict different equation-of-state parameter $w$ of the
dark energy and thus different cosmology. Another way to
distinguish the nature of dark energy is to measure its sound
speed $c_s$ which affects the perturbations in the energy
distribution and thus detectable through the observation of cosmic
microwave background(CMB) and large scale structure(LSS)
\cite{steinhardt2}. For cosmological constant and quintessence
with canonical Lagrangian, the sound speed is equal to unity(the
speed of light). But the sound speed of dark energy with
non-canonical Lagrangian can differ from unity and vary with time.

It is worth noting that recent observations do not exclude, but in
fact suggest a value even less than $-1$, indeed, they can lie in
the range $-1.38<w<-0.82$ \cite{melchiorri}. Especially, the new
results from SN-Ia alone are suggesting $w<-1$ at 1 $\sigma$
\cite{tonry}. So far, no fundamental justification is given for
the negative kinetic (phantom energy) term whose use is hence
motivated by the phenomenology. Phantom energy, a term coined by
Caldwell \cite{RRC} for matter with $w<-1$, certainly has some
strange properties. For example, its energy density increases with
time. Phantom field also violates the dominant-energy condition
\cite{caldwell} that might allow the existence of astrophysical or
cosmological wormholes and a striking consequence that the
universe will undergo a catastrophic "big rip" in a finite cosmic
time. Big rip corresponds to a new type of evolution of the
universe in which the expansion is so violent that all galaxies,
stars, planets and even atomic nuclei will be ripped apart in
finite cosmic time. However, it is avoidable that the universe
will undergo a catastrophic big rip \cite{hao9}. Matter with
$w<-1$ has received increased attention among theorists,  some
phantom models that possess negative kinetic energy are
investigated by many authors \cite{caldwell,caldwell1}. Recently,
Hao and Li \cite{Hao} proposed an interesting model of phantom
with a Born-Infeld (BI) type Lagrangian and show that current
universe is not a stable stage in such a model while is in its way
to the stable stage, at which the universe is dominated by the
vacuum energy like dark energy with a equation of state of $-1$.
BI phantom model was also studied in Ref. \cite{Nojiri}.

On the other hand, there are intensive interest in the brane world
scenario during the past several years\cite{brane12}. In contrast
to the conventional Kaluza-Klein (KK) picture where the
extra-dimensions are compacted on a small enough radius to evade
detection in the form of KK modes, the extra dimensions could be
large in the brane world scenario. Particles in the standard model
are expected to be confined to the brane, whereas the gravitons
propagate in the entire bulk spacetime, which gives an interesting
feature in the brane world scenario. In the Randall-Sundrum (RS)
second model \cite{Randall}, 4D Newtonian gravity is recovered at
low energies, because gravity is confined in a single
positive-tension brane even if the extra dimension is not compact.
For this model, many authors discussed the geometrical aspects
 \cite{Deffayet} as well as cosmology \cite{Deffayet2}. The purpose
of this work is to study the evolution of a BI type dark energy
with $w<-1$ on the brane world. We shall firstly discuss the
general property of the BI phantom, then analyze a concrete model
numerically.

We consider the second RS brand world scenario on a 3-brane, in
which the bulk is invariant under $Z_2$ reflection along the brane
and described by 5-dimensional Einstein gravity with a negative
cosmological constant and the brane motion is described by
Israel's junction condition. In this scenario, the bulk geometry
is AdS-Schwarzschild spacetime and the evolution of the brane is
governed by the following effective Friedmann equation:

\begin{equation}\label{H}
    H^2=\bigg(\frac{\dot{a}}{a}\bigg)^2=\frac{\kappa^4_5}{36}\rho_b^2-\frac{K}{a^2}+\frac{\mu}{a^4}-\frac{1}{l^2},
\end{equation}

\noindent where $\kappa_5$ is the 5-dimensional gravitational
coupling constant, $\rho_b$ is the energy density on the brane,
$K$ is the curvature constant with positive, zero and negative
value corresponding to spherical, flat and hyperbolic brane,
respectively, $l = \sqrt{\frac{6}{|\Lambda_5|}}$ is the length
scale of the bulk (negative) cosmological constant $\Lambda_5$,
and $\mu$ is related to the mass parameter of the bulk black hole.
The term $\frac{\mu}{a^4}$ is due to Weyl tensor in the bulk and
can be understood as dark radiation. This equation relates the
Hubble parameter to the energy density but it is different from
the usual Fridemann equation. The most remarkable feature of
Eq.(\ref{H}) is that the first term in the right hand side of this
equation is proportional to $\rho_b^2$ which is different from
that of the standard Friedmann equation which the energy density
enters linearly.   If one considers a brane with the total energy
density

\begin{equation}\label{rhob}
\rho_b=\rho+\sigma,
\end{equation}

\noindent where $\sigma$ is the tension of the brane which is
constant in time, and $\rho$ the energy density of ordinary
cosmological matter, then one obtains

\begin{equation}\label{H2}
    H^2=\frac{\kappa^2}{3}\rho\left(1+\frac{\rho}{\sigma}\right)+\frac{\mu}{a^4}-\frac{K}{a^2},
\end{equation}

\noindent where $\kappa^2 \equiv 8\pi G=
\frac{\kappa_5^4}{6}\sigma= \kappa_5^2l^{-1}$. It is clear that
the standard cosmology is recovered at low energy if the dark
radiation is neglected.

The BI phantom with Lagrangian as following \cite{Hao}:

\begin{equation}\label{lagoftp}
L=-V(\phi)
\sqrt{1-g^{\mu\nu}\partial_{\mu}\phi\partial_{\nu}\phi}.
\end{equation}

\noindent where $g_{\mu\nu}$ is the induced brane metric and
$V(\phi)$ is the potential of the model. On the homogeneous and
isotropic brane, we can rewrite the above Lagrangian as

\begin{equation}\label{Lagrangian2}
L=-V(\phi)\sqrt{1+\dot{\phi}^2}
\end{equation}

\noindent for the spatially homogeneous phantom field. The
equation of motion reads

\begin{equation}\label{ddPhi}
\ddot{\phi}
+3H\dot{\phi}(1+\dot{\phi}^2)-\frac{V^{'}(\phi)}{V(\phi)}(1+\dot{\phi}^2)=0
\end{equation}

\noindent where $H$ is the Hubble parameter as that in
Eq.(\ref{H}), the over dot represents the differentiation with
respect to $t$ and the prime denotes the differentiation with
respect to $\phi$. The density $\rho_{\phi}$ and the pressure
$p_{\phi}$ are defined as following:

\begin{equation}
\rho_{\phi}= \frac{V(\phi)}{\sqrt{1+\dot{\phi}^2}},
\end{equation}

\begin{equation}
p_{\phi}=-V(\phi)\sqrt{1+\dot{\phi}^2}.
\end{equation}

\noindent Therefore, the equation of state can be written as

\begin{equation}
w\equiv \frac{p_{\phi}}{\rho_{\phi}}=- \dot{\phi}^2-1,
\end{equation}

\noindent and the sound  speed

\begin{equation}
c_{s}^2 \equiv \frac{p_{\phi, X}}{\rho_{\phi, X}}=-w,
\end{equation}

\noindent where $X=\frac{1}{2}\dot{\phi}^2$. It is obvious that
the equation of state $w$ will be less than $-1$ and $c_s^2$
greater than the speed of light(unity), unless the kinetic energy
term $-\dot{\phi}^2=0$ .

Before discussing the property of general model, let us first
consider a simple model that $V=V_0$ where $V_0$ is a positive
constant. In this case, BI phantom behaves as Chaplygin gas with
$p_{\phi}=-\frac{V_0^2}{\rho_{\phi}}$ which is now a popular
candidate for dark energy. Meanwhile, one can find the obviously
difference with respect to the 4-dimensional case considered in
Ref.\cite{Hao}. There is a potentially observable difference
between the two scenarios for the specific potential. In the
specific case of $V_0>>\sigma$, the evolution of the brane is
determined by

\begin{equation}\label{11}
\frac{da}{adt}=\frac{\kappa V_0}{\sqrt{3\sigma (1+\dot{\phi}^2)}},
\end{equation}

\begin{equation}\label{12}
\frac{d(\dot{\phi}^2)}{dt}=\frac{\kappa
V_0}{\sqrt{3\sigma}}\dot{\phi}^2\sqrt{1+\dot{\phi}^2}.
\end{equation}

From Eqs.(\ref{11}) and (\ref{12}), we obtain the solution of the
scale factor that

\begin{equation}\label{13}
\frac{\sqrt{c_0}a^3+\sqrt{c_0a^6-1}}{\sqrt{c_0}a^3_0+\sqrt{c_0a^6_0-1}}=\exp\left[\sqrt{\frac{3}{\sigma}}\kappa
V_0(t-t_0)\right],
\end{equation}

\noindent where $c_0$, $a_0$ and $t_0$ are three positive
constants, and the equation-of-state parameter
$w=-\frac{1}{c_0a^6-1}-1<-1$. As above mentioned, it has been
pointed out by some authors that in the $w<-1$ case the fate of
the universe may be a big rip \cite{caldwell}. However, we notice
that $a\rightarrow \infty$ when and only when $t\rightarrow
\infty$, and there is, therefore, no such a doomsday for this
solution.

In the 4-dimensional case \cite{Hao} with the constant potential
which corresponds to the limit case $V_0<<\sigma$ in brane
scenario, we have a solution as follows

\begin{equation}
t-t_0=\frac{2}{3}a^{3/2}c_0^{1/4}F[\frac{1}{4},\frac{1}{4},\frac{5}{4};-\frac{a^6V_0^2}{c_0}],
\end{equation}

\noindent where $F$ denotes Hypergeometric Function, $c_0$ and
$t_0$ are two integral constants. It is easy to find that $a \sim
t^{2/3}$ for early time and $a\sim e^{\sqrt{\frac{V_0}{c_0}}t}$
for late time. But in Eq.(\ref{13}), $a(t)$ expands exponentially
not only in the late time but also in the early time. Therefore,
in the brane scenario, the universe comes into the accelerating
phase earlier, which is a potentially observable effect, and
$\sigma$ is an impressible and potentially observable parameter.
Furthermore, this argument can be generalized to the model with
other reasonable potentials.

 In general case, according to the equation of energy
conservation, we have

\begin{equation}
\dot{\rho_{\phi}}=\sqrt{3}\kappa\bigg(\rho_{\phi}+\frac{\rho_{\phi}^2}{\sigma}\bigg)^{1/2}\left(\frac{V_0^2}{\rho_{\phi}}-\rho_{\phi}\right).
\end{equation}

\noindent Because $\rho_{\phi}$ is always not greater than $V_0$,
the energy of BI phantom will grow increasingly until
$\rho_{\phi}$ is equal to $V_0$( at that time $\dot{\phi}=0$ and
hence $w$ becomes $-1$).

In a more realistic model we consider, including pressureless dust
case, the conservation equation and Friedmann equation can be
written as

\begin{equation}
\rho_{\phi}'+\rho_d'=3\left(\frac{V_0^2}{\rho_{\phi}}-\rho_{\phi}
- \rho_d\right),
\end{equation}

\begin{equation}
\dot{N}^2=\frac{\kappa^2}{3}[\rho_d+\rho_{\phi}+(\rho_d+\rho_{\phi})^2\sigma^{-1}],
\end{equation}

\noindent where $N\equiv\ln a$ and dot and prime denote
derivatives with respect to $t$ and $N$, respectively. We assume
that there is no interaction between dust and phantom,

\begin{equation}
\rho_d=\rho_{di}e^{3(N_i-N)},
\end{equation}

\begin{equation}
\rho_{\phi}=[V_0^2-(V_0^2-\rho_{\phi i}^2)e^{6(N_i-N)}]^{1/2}.
\end{equation}

\noindent where $N_i$, $\rho_{di}$ and $\rho_{\phi i}$ denote the
value of $N$, energy densities of dust and phantom at $t_i$
respectively. Therefore, with the growth of the scale factor, the
cosmic density parameter for dust
$\Omega_d\equiv\frac{\rho_d}{\rho_{tot}}$ decreases to zero, while
the cosmic density for BI phantom
$\Omega\equiv\frac{\rho_{\phi}}{\rho_{tot}}$ increases to unity,
where $\rho_{tot}$ denotes the total energy density on the brane.

Now, we turn to BI phantom with general potential on the brane.
For simplicity, we consider the spatially flat brane filled with
BI phantom, the energy density of non-relativistic matter and
radiation can be neglected, i.e. $\rho \simeq \rho_{\phi}$. In
this case, omitting the effect of dark radiation
($\frac{\mu}{a^4}$ term), we have

\begin{eqnarray}\label{maseq2}
\ddot{\phi}
&+&\sqrt{3}\kappa\dot{\phi}(1+\dot{\phi}^2)^{3/4}V^{1/2}(\phi)\left(1+
\frac{V(\phi)}{\sqrt{1+\dot{\phi}^2}}\sigma^{-1}\right)^{1/2}\nonumber\\
&-&\frac{V^{'}(\phi)}{V(\phi)}(1+\dot{\phi}^2)=0.
\end{eqnarray}

Introducing the new variables
\begin{equation}\label{newvar}
x=\phi, \hspace{0.5cm} y=\dot{\phi},
\end{equation}

\noindent then Eq.(\ref{maseq2}) becomes

\begin{eqnarray}\label{auto}
\frac{d x}{dt}&&=y,\\\nonumber \frac{d y}{dt}&&=(1+y^2)
\frac{V'(x)}{V(x)}\\\nonumber&&-\sqrt{3}\kappa y(1+y^2)^{3/4}
V(x)^{1/2}[1+\sigma^{-1}V(x)(1+y^2)^{-1/2}]^{1/2}.
\end{eqnarray}

\noindent Linearizing the above system around its critical point
$(x_c, 0)$ where the value of $x_c$ is determined by $V'(x_c)=0$,
one obtain the following linear system

\begin{eqnarray}\label{linear}
\frac{dx}{dt}&&= y,\\\nonumber
\frac{dy}{dt}&&=\frac{V''_c}{V_c}x-\kappa\sqrt{3V_c(1+\sigma^{-1}
V_c)} y,
\end{eqnarray}

\noindent where $V_c$ is the value of potential at the critical
point which is a stable node when $V_c''<0$ and a saddle when
$V_c''>0$. Therefore, the system permits attractor solutions when
the potential of the Born-Infeld scalar field have a maximum.

For the  brane with strong tension, $\sigma >> V_0$, the
Eq.(\ref{maseq2}) recovers the case of standard Einstein frame
which have been considered in Ref.\cite{Hao}, and for the brane
with weak tension, $\sigma << M_p^4$, we have

\begin{equation}\label{maseq3}
\ddot{\phi}
+\frac{\sqrt{3}\kappa}{\sqrt{\sigma}}V(\phi)\dot{\phi}\sqrt{1+\dot{\phi}^2}-\frac{V'}{V}(1+\dot{\phi}^2)
=0.
\end{equation}

\noindent Especially, we choose a potential
$V(\phi)=\frac{2\sqrt{\sigma}}{\sqrt{3}\kappa}(1+\phi^2)^{-1/2}$
with an maximum at $\phi=0$. There exists an exact solution
$\phi\sim e^{-t}$, which implies that phantom will decay into a
cosmological constant exponentially, meanwhile the scale factor of
the brane goes with $a\sim (1+e^{2t})^{1/2}$. We notice that
$a\rightarrow \infty$ when and only when $t\rightarrow \infty$,
and hence there is also no 'big rip' for this solution.

Next, we consider another potential as follows

\begin{equation}\label{potential}
    V(\phi)=V_0(1+\frac{\phi}{\phi_0}) e^{-\phi/\phi_0}
\end{equation}

\noindent which is used to be describe the rolling tachyon
\cite{kutasov}. Clearly, this potential have also a maximum value
$V_0$ at $\phi=0$. By rescaling, Eq.(\ref{maseq2}) can be reduced

\begin{eqnarray}\label{eqnddphi}
\ddot{\phi}
&+&\alpha\dot{\phi}(1+\dot{\phi}^2)^{3/4}(1+\phi)^{1/2}e^{-\phi/2}\left(1+
\beta\frac{(1+\phi)e^{-\phi}}{\sqrt{1+\dot{\phi}^2}}\right)^{1/2}\nonumber\\
&+&\frac{\phi}{1+\phi}(1+\dot{\phi}^2)=0,
\end{eqnarray}

\noindent where the dimensionless parameters
$\alpha=\sqrt{3V_0}\kappa\phi_0$ and $\beta=\frac{V_0}{\sigma}$,
respectively.

\begin{figure}
\epsfig{file=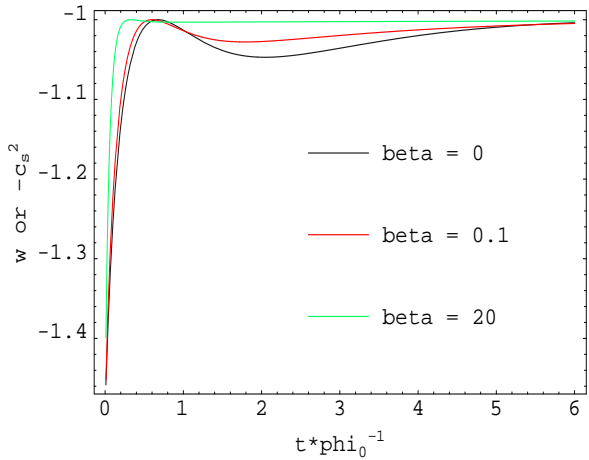,height=2.5in,width=3.2in} \caption{The
evolution of the equation-of-state parameter $w$ of phantom with
respect to $t$ where the parameter $\alpha=2$ and $\beta=0$,
$0.1$, $20$, respectively.}
\end{figure}

The numerical results with different brane tension are shown in
FIG.1, from which we can find that the BI phantom behaves as a
cosmological constant at the late time ($c^2_s\rightarrow 1 $ and
$w\rightarrow -1$ when $t\rightarrow\infty$) and the weaker the
brane tension is, the faster the BI phantom comes into
cosmological constant.

When matter and radiation are also considered, we have

\begin{eqnarray}
\frac{dx}{dN}&=& y
H_i^{-1}E(N)^{-1/2}[1+\frac{\beta}{\alpha}E(N)]^{-1/2},\nonumber\\
\frac{dy}{dN}&=&-3y(1+y^2)\nonumber\\
&-&\frac{x(1+y^2)}{1+x}H_i^{-1}E(N)^{-1/2}[1+\frac{\beta}{\alpha}E(N)]^{-1/2},
\end{eqnarray}

\noindent where
$E(N)=\Omega_{m,i}e^{-3N}+\Omega_{r,i}e^{-4N}+\alpha\frac{(1+x)e^{-x}}{\sqrt{1+y^2}}$,
$H_i$, $\Omega_{m,i}$ and $\Omega_{r,i}$ are Hubble parameter,
density parameter of matter and radiation at $t_i$, respectively.
The parameter $\alpha = \frac{V_0}{\rho_{c,i}}$ where $\rho_{c,i}$
is the critical density at $t_i$. The FIG.2 plots the evolution of
density parameter of BI phantom $\Omega_{\phi}$ with the different
value of $\beta=\frac{V_0}{\sigma}$. We find again that the weaker
the brane tension is, the earlier the BI phantom becomes dominant,
which is potentially observable effect.

\begin{figure}
\epsfig{file=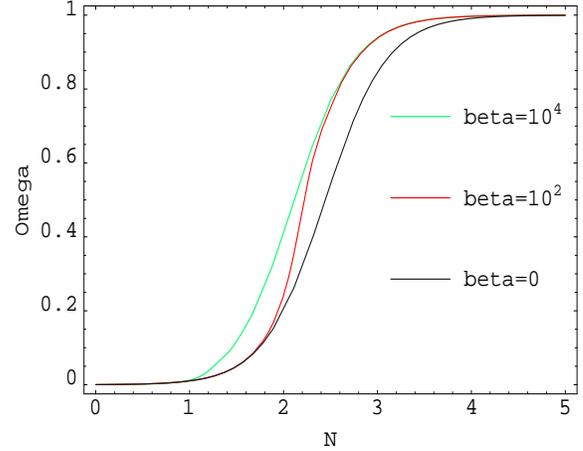,height=2.5in,width=3.2in} \caption{The
evolution of the density parameter of BI phantom with respect to
$N$.}
\end{figure}

 In summary, the behavior of the BI phantom on the brane will
attract to a cosmological constant for the model with a potential
having a maximum and the brane will be dominated by the BI phantom
at the late time. Even though these models are all asymptotically
de Sitter, in the sense that they are indistinguishable from dS
space at very late time, the tension of the brane $\sigma$ is a
potentially observable quantity which determines sensitively the
evolution of the scale factor $a$. We obtain an exact solution for
the BI phantom and the scale factor of the brane for a model
satisfying above condition in weak tension limit and find that the
brane will not trend to a big rip. It is of importance to point
out that the BI phantom, which can be treated as a realization of
generalized Chaplygin gas model for dark energy, may exist not
only nowadays, but also in the very early period of the universe
while the effect of the brane might be leading.

\vspace{0.8cm} \noindent ACKNOWLEDGEMENT: This work was partially
supported by National Nature Science Foundation of China under
Grant No. 19875016, and Foundation of Shanghai Development for
Science and Technology under Grant No.01JC14035.


\begin{thebibliography}{99}
\bibitem{newobservation} P. de Bernardis et al., Nature {\bf 404} 955 (2000); S. Hanany et al.
Astrophys. J. {\bf 545} 1 (2000); N. Bahcall, J. P. Ostriker, S.
Perlmutter and P. J. Steinhardt, Science {\bf 284} 1481 (1999); S.
Perlmutter et al., Astrophys. J. {\bf 517} 565 (1999 ); A. G.
Riess et al., Astron. J. {\bf 116} 1009 (1998).

\bibitem{WMAP} C. Bennett \textit{et al}., astro-ph/0302207; G.
Hinshow \textit{et al}., astro-ph/0302217; A Kogut et al.,
astro-ph/0302021.
\bibitem {steinhardt} B. Ratra and P. J. E. Peebles, Phys. Rev. {\bf D37} 3406 (1988);
R. R. Caldwell, R. Dave and P. J. Steinhardt, Phys. Rev. Lett.
{\bf 80} 1582 (1998); P. J. Steinhardt, L. Wang and I. Zlatev,
Phys. Rev. {\bf D59} 123504 (1999); I. Zlatev, L. Wang and P. J.
Steinhardt, Phys. Rev. Lett. {\bf 82} 896 (1999); K. Coble, S.
Dodelson, J. A. Frieman, Phys. Rev. {\bf D55} 1851 (1997); X. Z.
Li, J. G. Hao, D. J. Liu, Class. Quantum Grav. \textbf{19} 6049
(2002); X. Z. Li, D. J. Liu and J. G. Hao, Chin. Phys. Lett.
\textbf{19}, 1584 (2002).

\bibitem{steinhardt2} J. K. Erickson \textit{et al}, Phys. Rev. Lett. \textbf{ 88} 121301 (2002);
 S. DeDeo, R. R. Caldwell and P. J. Steinhardt, Phys. Rev.
\textbf{D67}, 103509 (2003).

\bibitem{melchiorri} A. Melchiorri, L. Mersini, C. J.
Odmann and M. Trodden, Astro-ph/0211522.
\bibitem{tonry} J. L. Tonry \textit{et al.}, astro-ph/0305008.
\bibitem{RRC}R. R.
Caldwell, Phys. Lett. \textbf{B545}, 23 (2002).

\bibitem{caldwell}R. R. Caldwell, M. Kamionkowski and N. N.
Weinberg, astro-ph/0302506; G. W. Gibbons, hep-th/0302199.

\bibitem{hao9}J. G. Hao and X. Z. Li, hep-th/0306033.

\bibitem {caldwell1}V. Sahni and A. A. Starobinsky, Int. J. Mod. Phys.
\textbf{D9} 373 (2002); L. Parker and A. Raval, Phys. Rev.
\textbf{D60} 063512 (1999); T. Chiba, T. Okabe and M. Yamaguchi,
Phys. Rev. \textbf{D62} 023511 (2000); B. Boisseau, G.
Esposito-Farese, D. Polarski and A. A. Starobinsky, Phys. Rev.
Lett. \textbf{85} 2236 (2000); A. E. Schulz, Martin White, Phys.
Rev. \textbf{D64} 043514 (2001); V. Faraoni, Int. J. Mod. Phys.
\textbf{D11} 471 (2002); I. Maor, R. Brustein, J. McMahon and P.
J. Steinhardt, Phys. Rev. \textbf{D65} 123003 (2002); V. K. Onemli
and R. P. Woodard, Class. Quant. Grav. \textbf{19} 4607 (2002); D.
F. Torres, Phys. Rev. \textbf{D66} 043522 (2002); S. M. Carroll,
M. Hoffman, M. Trodden, astro-ph/0301273; P. H. Frampton,
hep-th/0302007; J. G. Hao and X. Z. Li, Phys. Rev. \textbf{D67},
107303 (2003); X. Z. Li and J. G. Hao, hep-th/0303093; A.
Feinstein and S. Jhingan,
 hep-th/0304069; L. P. Chimento and A. Feinstein,
astro-ph/0305007; P. Singh, M. Sami and N. Dadhich,
hep-th/0305110; S. Nojiri and S. D. Odintsov, Phys. Lett.
\textbf{B562} 147 (2003), hep-th/0303117; S. Nojiri and S. D.
Odintsov, Phys. Lett. \textbf{B565} 1 (2003), hep-th/0304131.

\bibitem{Hao} J. G. Hao and X. Z. Li, Phys. Rev. \textbf{D68}, 0435XX (2003), in press,
hep-th/0305207.
\bibitem{Nojiri} S. Nojiri and S. D. Odintsov, hep-th/0306212.
\bibitem{brane12}N. Arkani-Hamed, S. Dimopoulos and G. Dvali,
Phys. Lett. \textbf{B429}, 263 (1998); I. Antoniadis, N.
Arkani-Hamed, S. Dimopoulos and G. Dvali, Phys. Lett.
\textbf{B436}, 257 (1998); N. Arkani-Hamed, S. Dimopoulos and G.
Dvali, Phys. Rev. \textbf{D59}, 086004 (1999); T. Shiromizu, K.
Maeda and M. Sasaki, Phys. Rev. \textbf{D62}, 024012 (2000); G.
Dvali, G. Gabadadze, M. Kolanovic and F. Nitti, Phys. Rev.
\textbf{D64}, 084004 (2001).

\bibitem{Randall}L. Randall and R. Sundrum, Phys. Rev. Lett. \textbf{83},
3370 (1999); \textit{ibid}, \textbf{83}, 4690 (1999).

\bibitem{Deffayet}C. Deffayet, G. Dvali, G. Gabadadze and A.
Vainshtein, Phys. Rev. \textbf{D65}, 044026 (2002); C. Deffayet.
Phys. Rev. \textbf{D66}, 103504 (2002); G. Kofinas, E.
Papantonopoulos and I. Pappa, Phys. Rev. \textbf{D66}, 104014
(2002); G. Kofinas, E. Papantonopoulos and V. Zamarias, Phy. Rev.
\textbf{D66}, 104028 (2002).

\bibitem{Deffayet2} C. Deffayet, Phys. Lett. \textbf{B502}, 199 (2001); C.
Deffayet, G. Dvali and G. Gabadadze, Phys. Rev. \textbf{D65},
044023 (2002); C. Deffayet, S. J. Landau, J. Raux, M. Zaldarriaga
and P. Astier, Phys. Rev \textbf{D66}, 024019 (2002); N. J. Kim,
H. W. Lee and Y. S. Myung, Phys. Lett. \textbf{B504}, 323 (2001).

\bibitem{kutasov}A. A. Gerasimov and S.L. Shatashvili, \textit{J.High Energy Phys.} \textbf{0010},
034 (2000);
 D. Kutasov, M. Marino and G. W. Moore, \textit{J.High Energy Phys.} \textbf{0010}, 045 (2000).


\end{thebibliography}
\end{document}